\documentclass[12pt]{article}
\usepackage{epsfig}
\begin{document}
\begin{center}
{\Large {\bf Effective constraint potential in lattice Weinberg - Salam model}


{
\vspace{1cm}
{ M.I.~Polikarpov$^{a}$ }, { M.A.~Zubkov$^{a}$ }\\
\vspace{.5cm} {\it $^a$ ITEP, B.Cheremushkinskaya 25, Moscow, 117259, Russia }
}}

\end{center}

\begin{abstract}
We investigate lattice Weinberg - Salam model without fermions for the value of
the Weinberg angle $\theta_W \sim 30^o$, and bare fine structure constant
around $\alpha \sim \frac{1}{150}$. We consider the value of the scalar self
coupling corresponding to bare Higgs mass around $150$ GeV. The effective
constraint potential for the zero momentum scalar field is used in order to
investigate phenomena existing in the vicinity of the phase transition between
the physical Higgs phase and the unphysical symmetric phase of the lattice
model. This is the region of the phase diagram, where the continuum physics is
to be approached. We compare the above mentioned effective potential
(calculated in selected gauges) with the effective potential for the value of
the scalar field at a fixed space - time point.   We also calculate the
renormalized fine structure constant using the correlator of Polyakov lines and
compare it with the one - loop perturbative estimate.
\end{abstract}



\section{Introduction}

 In lattice Electroweak theory the importance of
the vicinity of the phase transition between the physical Higgs phase and the
unphysical symmetric phase is related to the fact that this is the region of
the phase diagram, where the continuum physics is to be approached. During the
early studies of lattice Weinberg - Salam model it was recognized that the
physics of the mentioned transition is intimately related to the way the
continuum physics arises within the lattice model. Namely, in \cite{MontvayPT}
it was suggested that the appearance of the second order phase transition leads
to the conventional picture: It is possible to increase infinitely the
ultraviolet cutoff
 along the line of constant physics corresponding to realistic values of
renormalized couplings. At the same time, according to \cite{MontvayPT} the
first order phase transition would  lead to another picture: the line of
constant physics intersects the phase transition line at a certain value of the
Ultraviolet cutoff $\Lambda_c$ that is in this case the maximal possible value
of the cutoff in the theory for the given values of renormalized couplings. It
is worth mentioning that the first order phase transition was shown to take
place at unphysical values of couplings \cite{MontvayPT}. However, at physical
values of coupling constants two state signal was not found that means that we
may deal either with the weak first order phase transition or with the second
order phase transition.

Yet another possibility was suggested in \cite{Z2010}: the transition might
appear to be a crossover. Starting from the physical Higgs phase and moving
towards the symmetric phase one observes the increase of Nambu monopole
\cite{Nambu,Chernodub_Nambu} density. (These objects are, in essence, the
embryos of the symmetric phase within the Higgs phase.) At a certain point on
the phase diagram  the average distance between these embryos becomes of the
order of their size. Further movement towards the symmetric phase leads to the
point, where the only minimum of the ultraviolet effective constraint potential
is at $\phi = 0$ while within the Higgs phase it has minimum at nonzero $\phi$
(see the next section for the definition of the potential). At this point,
however, the $Z$ - boson mass as well as the Higgs boson mass does not vanish
(both masses are defined in the Unitary gauge when the scalar field is real and
not negative). Then one can move further with the increase of the cutoff, and
there should exist the point, where the transition to the true symmetric phase
occurs (in this phase the gauge boson masses must vanish). According to
\cite{Z2010} the vicinity of the phase transition, where the Nambu monopoles
dominate, is called the fluctuational region. The possibility to describe the
continuum Electroweak physics within the fluctuational region is questionable
due to the Nambu monopoles that are supposed to give unexpected contributions
to the physical observables. In particular, the transition might be a crossover
(while the tree level perturbative effective action predicts the second order
phase transition).

 We would like to suggest a further analogy with the
superconductor theory. Namely, for the second order superconductors in the
presence of the external magnetic field there exist several pseudocritical
lines on the phase diagram. First, when the value of the magnetic field
achieves the value $H_{c1}$ the Abrikosov vortices are formed. These objects
are the embryos of the normal phase within the superconducting one. The mixed
phase is formed, where the lattice of Abrikosov vortices exists within the
superconductor. Next, when the value of magnetic field achieves the
pseudocritical value $H_{c2}$, the mixed phase is transformed to the normal
phase. We suppose that the mixed phase of the second order superconductors is
similar to the fluctuational region of the lattice Electroweak theory mentioned
above. Of course, in our case we do not have any external field and the $Z$ -
vortices and Nambu monopoles arise spontaneously. Therefore, it is necessary to
take care when applying the given analogy.

It is worth mentioning that the effective Abelian gauge model appears within
the Weinberg - Salam model with the $Z$ - boson playing the role of the Abelian
gauge field. The second order superconductor appears within the Ginzburg -
Landau model for $M_H
> M_Z$. Therefore for $M_H > M_Z$ the Weinberg - Salam model may be similar to
the second order superconductors (and not to the first order superconductors).
For this reason, we do not expect, in particular, the appearance of the first
order phase transition for $M_H > M_Z$. (The first order phase transition takes
place for the first order superconductors.)

In the present paper we proceed with the research of \cite{Z2010} and consider
the fluctuational region more carefully. The measurements were performed at
much more  different points in the vicinity of the transition than it was done
in \cite{Z2010}. However, as a price for this we simulate the system on smaller
lattices. Our main results reported here are obtained on the lattice $8^3\times
16$ while in \cite{Z2010} the lattices up to the size $20^3\times 24$ were
used. As an additional device for the investigation we use the effective
potential for the zero - momentum scalar field. In order to consider such a
potential the gauge is to be fixed. We consider two different ways to fix the
gauge and investigate the resulting effective potentials. It is shown, that one
of the given potentials changes its form at the point $\gamma_c$, where the
mentioned above ultraviolet effective potential changes its form. The other
potential changes its form at the value $\gamma_c^{\prime}$ different from
$\gamma_c$. This contradicts with the conventional picture that is based on the
perturbation theory and implies that the scalar field condensates defined in
different gauges vanish at the same point of the phase diagram. Thus the
hypothesis that the given transition is a crossover is confirmed (at least on
the lattices of considered sizes): different quantities change their behavior
at different points on the phase diagram. However, the perturbation theory does
not appear to be completely useless in the vicinity of the transition. Namely,
we calculate the renormalized fine structure constant using the methods
different from that of \cite{Z2010} and obtain a surprising coincidence with
the 1 - loop estimate.

 The paper is organized as
follows. In Section 2 we consider the definition of  the  effective potentials
under consideration. In Section 3 we consider the details of lattice
regularized Weinberg - Salam model. In Section 4 we list our numerical results.
In section 5 we discuss the obtained numerical results. Throughout the paper
the notations of differential forms on the lattice are used (for their
definition see, for example, \cite{forms}).

\section{Infrared effective potential}

In ordinary lattice scalar field theory (the real  scalar field $h_x$ is
defined on the lattice points $x$) there exist several definitions of effective
constraint potentials. Namely, one may consider the ultraviolet potential

\begin{equation}
exp(-V^{u-v}(\phi)) = <\delta(\phi - h_x)>\label{uv}
\end{equation}

Also it is possible to consider the infrared potential
\begin{equation}
exp(-V^{i-r}(\phi) )= <\delta(\phi - |\frac{1}{N} \sum_x h_x|)>, \label{ir}
\end{equation}
where $N$ is the number of lattice points.

In principle, it is expected that the given potentials have nontrivial minima
at nonzero $\phi$ in the broken phase, where the scalar field is condensed.
However, in a more complicated model this statement is questionable because the
value $\phi_m$, at which the infrared potential has its minimum, is an infrared
quantity while potential (\ref{uv}) is at a first look an ultraviolet quantity.

In the lattice theory for the complex scalar field charged with respect to
lattice $U(1)$ gauge field $Z\in (-\pi;\pi]$ there are several complications.
After fixing the gauge $H = \left(\begin{array}{c}h\\0\end{array}\right), h \in
C$, where $H$ is the scalar doublet, the lattice Weinberg - Salam model becomes
such a lattice gauge - Higgs model with the complex  scalar field charged with
respect to the $Z$ - boson that plays now the role of the $U(1)$ gauge field.

Now (\ref{ir}) is not gauge invariant and, therefore, has the only minimum at
$\phi = 0$ everywhere. There exist also naive gauge invariant version of the
infrared potential:
\begin{equation}
exp(-V^{\prime}_{i-r}(\phi)) = <\delta(\phi - \frac{1}{N}\sum_x
|h_x|)>,\label{irmod}
\end{equation}
However, this potential, obviously, has the only minimum at $\phi \ne 0$ in
both phases.

Instead of (\ref{irmod}) we can consider  potential (\ref{ir}) at a fixed
gauge. Actually, (\ref{irmod}) is equivalent to (\ref{ir}) for the version of
the Unitary gauge adopted in \cite{Z2010}: $h_x \in R, h_x \ge 0$.

If the Unitary gauge is fixed using only the condition $h_x \in R$, the $Z_2$
gauge degrees of freedom remain: $h_x \rightarrow (-1)^{n_x} h_x, \, Z
\rightarrow [Z + \pi d n]{\rm mod}\, 2 \pi$. This remaining gauge freedom is to
be the subject of the further gauge fixing. The simplest choice here is
$h_x>0$. However, this choice does not lead to the effective potential
sensitive to the transition between the two phases.

The other possible choice is minimization of
\begin{equation}
\sum_{links}(1-{\rm cos}\, Z) \rightarrow min \label{Z_}
\end{equation}
 with respect to the mentioned $Z_2$
transformations. Further we refer to this gauge as to the $Z$ - version of
Unitary gauge and refer to the corresponding effective potential (\ref{ir}) as
to UZ potential.

Yet another way to define the Unitary gauge with $h_x \in R$ is to minimize the
divergence of $Z$ with respect to the remaining $Z_2$ transformations:
\begin{equation}
\sum_x [\delta Z]^2 \rightarrow min \label{dz}
\end{equation}
Further we refer to this gauge as to the $DZ$ - version of Unitary gauge and
refer to the corresponding effective potential (\ref{ir}) as to UDZ potential.
It will be shown that the infrared potential in this gauge has the only minimum
at $\phi = 0$ in the region of the phase diagram, where potential (\ref{uv})
also has the only minimum at $\phi = 0$. In the region of the phase diagram,
where the ultraviolet potential has its minimum at nonzero $\phi$, the infrared
potential (\ref{ir}) has its minimum at nonzero $\phi$ as well.

In principle, there exists also the possibility to consider the infrared
effective potential for the scalar field surrounded by the Z - boson cloud (the
Dirac construction \cite{Dirac}) that is gauge invariant by definition.
However, the corresponding numerical procedure is rather time consuming,
especially on large lattices. Therefore, at the present moment we do not
consider such a construction.

\section{The lattice model under investigation}

We consider lattice Weinberg - Salam model without fermions. The partition
function has the form:

\begin{equation}
Z = \int D H D\Gamma exp(-A(\Gamma,H))
\end{equation}

Here $A(\Gamma,H)$ is the action for the scalar doublet $H$ and the gauge field
$\Gamma = U\otimes e^{i\theta} \in SU(2)\otimes U(1)$:
\begin{eqnarray}
 A(\Gamma,H) & = & \beta \!\! \sum_{\rm plaquettes}\!\!
 ((1-\mbox{${\small \frac{1}{2}}$} \, {\rm Tr}\, U_p )
 + \frac{1}{{\rm tg}^2 \theta_W} (1-\cos \theta_p))+\nonumber\\
 && - \gamma \sum_{xy} Re(H_x^+U_{xy} e^{i\theta_{xy}}H_y) + \sum_x (|H_x|^2 +
 \lambda(|H_x|^2-1)^2), \label{S}
\end{eqnarray}

The action can be rewritten as follows:
\begin{eqnarray}
 A(\Gamma,H) & = & \beta \!\! \sum_{\rm plaquettes}\!\!
 ((1-\mbox{${\small \frac{1}{2}}$} \, {\rm Tr}\, U_p )
 + \frac{1}{{\rm tg}^2 \theta_W} (1-\cos \theta_p))+\nonumber\\
 && + \frac{\gamma}{2} \sum_{xy} |H_x - U_{xy} e^{i\theta_{xy}}H_y|^2 + \sum_x (|H_x|^2 (1 - 2\lambda - 4\gamma) +
 \lambda |H_x|^4), \label{S}
\end{eqnarray}

Now we easily derive expressions for the tree level vacuum expectation value
$v$ of $|H_x|$, the lattice Higgs boson mass $m_H = M_H a$, the lattice $Z$ -
boson mass $m_Z = M_Z a$, and the critical value $\gamma_c$:
\begin{eqnarray}
v &=& \sqrt{2\frac{\gamma - \gamma_c}{\lambda}}\nonumber\\
m_H &=& v\sqrt{\frac{8\lambda}{\gamma}}\nonumber\\
m_Z &=& v\sqrt{\frac{\gamma}{\beta \, {\rm cos}^2 \theta_W}}\nonumber\\
\gamma^{(0)}_c &=& \frac{1 - 2\lambda}{4}\label{tree}
\end{eqnarray}
After fixing the gauge $H = \left(\begin{array}{c}h\\0\end{array}\right), h \in
C$, where $H$ is the scalar doublet, the lattice Weinberg - Salam model becomes
the lattice gauge - Higgs model with the scalar field charged with respect to
the $Z$ - boson that plays now the role of the $U(1)$ gauge field. Natural
definition of the $Z$ - boson field is $Z = {\rm Arg}\, [U_{11}e^{i\theta}]$.
Next, after fixing the Unitary gauge the field $h$ becomes real. However, the
$Z_2$ gauge ambiguity remains: $h_x \rightarrow (-1)^{n_x} h_x, \, Z
\rightarrow [Z + \pi d n]{\rm mod}\, 2 \pi$.

The tree level approximation gives for the effective constraint potential
(\ref{uv}):
\begin{equation}
V^{u-v}(\phi) =  -3 \, {\rm log} \phi + \frac{\gamma}{2G_{m_H}(0)} (\phi - v
)^2 \label{UVPOT}
\end{equation}

Here we encounter the lattice volume $N$ and the value of lattice Yukawa
potential at zero distance $ G_{m_H}(0) = \frac{1}{N} \sum_p (4 sin^2p/2 +
m_H^2)^{-1}$.  Even on the infinite lattice  this value remains finite if $m_H$
is nonzero: $G_{m_H}(0) < 1/m_H^2$. This means that the ultraviolet
fluctuations $\delta \phi = \sqrt{G_{m_H}(0)/\gamma}$ expressed in lattice
units remain finite. When the physical volume of the lattice $Na^4 >> M_H^{-4}$
is kept constant while the lattice spacing tends to zero and, consequently,
$m_H=M_H a \rightarrow 0$ (here $M_H$ is the Higgs mass in GeV, $a$ is the
lattice spacing) the value of $G_{m_H}(0)$ remains finite.  Thus we recover the
usual prediction of the continuum theory $\delta \phi^{phys} =
\sqrt{G^{phys}_{M_H}(0)/\gamma}\sim \Lambda = \frac{\pi}{a}$, where
$\phi^{phys}$ is the scalar field expressed in physical units while
$G^{phys}_{M_H}(x)$ is its propagator in physical units (here in lattice
regularization, though).

The tree level approximation gives for the infrared effective constraint
potential (\ref{ir}):
\begin{equation}
V^{i-r}(\phi) =  N \lambda (\phi^2 - v^2 )^2 \label{IRPOT}
\end{equation}

This expression corresponds also to mean field approximation. Here $v$ is the
same as for the ultraviolet potential. Unlike the ultraviolet effective
potential, however, the fluctuations of $\phi$ decrease fast with the increase
of the lattice size.
  When the physical volume $\cal V$ of the
lattice ${\cal V} = Na^4 $ is kept constant while the lattice spacing tends to
zero and, consequently, $m=M_H a \rightarrow 0$ (here $M_H$ is the Higgs mass
in GeV, $a$ is the lattice spacing) the fluctuations of $\phi$ in lattice units
tend to zero while fluctuations in physical units $\sim 1/\sqrt{8\lambda
v_{phys}^2 {\cal V}}$ remain finite. Thus in mean field approximation the
nontrivial minimum of the infrared constraint effective potential appears at
the same value  of the field as the nontrivial minimum of the ultraviolet
effective potential. Taking into account  loop corrections one would come, in
principle, to the expression for the infrared effective constraint potential in
the form of Coleman - Weinberg. (See, for example, \cite{LoopPT}, where the
finite temperature version of the potential was given.)

\section{Numerical results}

We investigated numerically the system at $\beta = 12$, $\lambda = 0.0025$,
$\theta_W = 30^o$. Our results were obtained mainly on the lattice $8^3\times
16$. According to the previous results \cite{Z2010} at these values of
couplings in the vicinity of the transition between the two phases the Higgs
boson mass is around $150$ GeV while bare fine structure constant is $\sim
1/150$. In \cite{Z2010} the transition point was localized at the point, where
the ultraviolet effective constraint potential (described in Section 2 of the
present paper) looses the nontrivial minimum at nonzero value of $\phi$.
Namely, at $\gamma > \gamma_c=0.26\pm0.001$ the value of $v$ calculated using
this type of effective potential is nonzero while for $\gamma \le \gamma_c$ the
value of $v$ vanishes. However, there are also two other selected points,
denoted in \cite{Z2010} by $\gamma_{c0}$ and $\gamma_{c2}$.

At $\gamma_{c0}$ the dependence of the lattice $Z$ - boson mass on $\gamma$
gives $m_Z = 0$. The linear fit to the Z - boson mass calculated in
\cite{Z2010}  indicates that $\gamma_{c0} = 0.252\pm 0.001$. However, we do not
exclude that the Z - boson mass may vanish at the values of $\gamma$ larger
than $0.252$. Actually, in \cite{Z2010} the nonzero Z - boson mass was obtained
for $\gamma\ge 0.258$.

At $\gamma_{c2} \sim 0.262$ the average distance between Nambu monopoles
becomes compared to their size. So, the fluctuational region is localized
between $\gamma_{c0}$ and $\gamma_{c2}$. Within this region it is expected that
the perturbation theory does not work well and nonperturbative phenomena become
important.

\subsection{Simulation details}

The model is simulated in Unitary gauge with the signs of $h$ unfixed.
Therefore, the $Z_2$ gauge freedom remains (together with the Electromagnetic
$U(1)$). In order to simulate the system the Metropolis algorithm is used. The
new suggested value of the gauge field is obtained via the (right)
multiplication of the old gauge field at the given link by the $SU(2)\times
U(1)$ random matrix. The values of this matrix are distributed randomly (with
Gauss distribution) around unity. Norm of the Gauss distribution is tuned
automatically in order to keep acceptance rate around $0.5$. The new suggested
value of the scalar field $h$ is obtained adding to the old one the value
$\delta h$ distributed randomly (with Gauss distribution) around zero. The norm
of this distribution (different from the norm used for the gauge fields) is
also tuned in order to keep acceptance rate around $0.5$. Each step of
Metropolis procedure contains suggestions of new fields at all points and links
of the lattice. The procedure starts from the zero values of the scalar fields
and the values of the gauge fields equal to unity (cold start). At each
considered value of $\gamma$ several independent processes run (up to $100$
processes). Out of the interval $\gamma \in [0.2575; 0.26]$ the equilibrium is
achieved after $30000$ Metropolis steps. (Far from the phase transition the
convergence is even faster.) The autocorrelation time for the gauge fields is
about $800$ Metropolis steps. For the scalar field the autocorrelation time is
one order of magnitude smaller. The code was tested in several ways. In
particular, some previous results in the $SU(2)$ gauge - Higgs model
\cite{MontvayPT,Montvayold,Montvay} and the Weinberg - Salam model with frozen
radius of the scalar field \cite{VZ2008} were reproduced.

\subsection{Cold start and Hot start}

\begin{figure}
\begin{center}
 \epsfig{figure=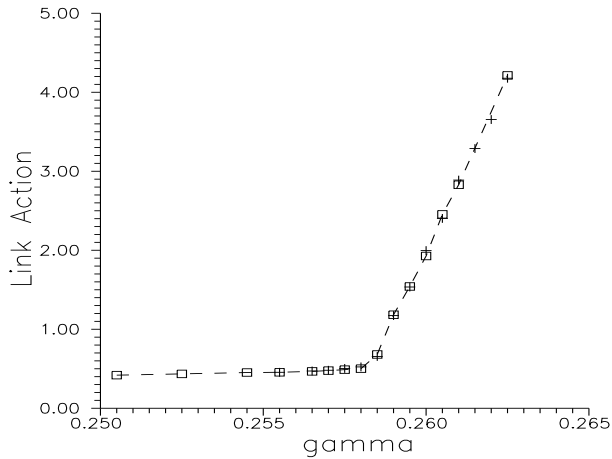,height=60mm,width=80mm,angle=0}
\caption{\label{LAQ} The link part of the action as a function of $\gamma$ at
$\lambda =0.0025$ , $\beta = 12$. Cold start corresponds to squares. Hot start
corresponds to crosses. The error bars are about of the same size as the
symbols used. }
\end{center}
\end{figure}

When the simulation starts from the cold start, the equilibrium is achieved
approximately after $30 000$ Metropolis steps for $\gamma > 0.26$ and $\gamma <
0.257$. (Far from the transition the number of steps needed in order to achieve
equilibrium is smaller.) This way we obtain, in particular, the equilibrium at
$\gamma = 0.255$. Next, starting from the corresponding configurations the
further simulation procedure is applied with the values of $\gamma$ between
$0.256$ and $0.264$. We call this simulation the hot start
simulation\footnote{The simulation that begins from the completely disordered
configurations converges to the equilibrium much more slow. This is because the
Algorithm has to overcome the confinement - deconfinement phase transition due
to $U(1)$ and to transfer the configuration with zero $\beta$ to the
configuration with rather high value $\beta = 12$. In practise such a
simulation never achieves equilibrium for the number of Metropolis steps up to
$200000$ on the lattice $8^3\times 16$. However, we observed the convergence on
smaller lattices of sizes up to $6^4$. }. The results of these simulations are
presented in Fig. \ref{LAQ}. In this figure the data of the link part of the
action $\frac{1}{4N}\sum_{xy}H_x^+U_{xy} e^{i\theta_{xy}}H_y$ (that is most
sensitive to the transition between the Higgs phase and the symmetric phase)
are represented. The squares correspond to the cold start ($450 000$ Metropolis
steps)\footnote{This simulation has required about 480 ours CPU time.}. The
crosses correspond to the hot start ($350 000$ Metropolis steps). One can see
that the two lines merge together. However, at earlier stages of the simulation
the Hysteresis pattern was observed that points to the interval $[0.257, 0.26]$
as to the place of the transition between the two phases. The convergence to
the equilibrium in this simulation is rather slow.  Our  data demonstrate the
absence of the two - state signal for $\gamma \in [0.257,0.26]$.

\subsection{Ultraviolet Effective potential}

\begin{figure}
\begin{center}
 \epsfig{figure=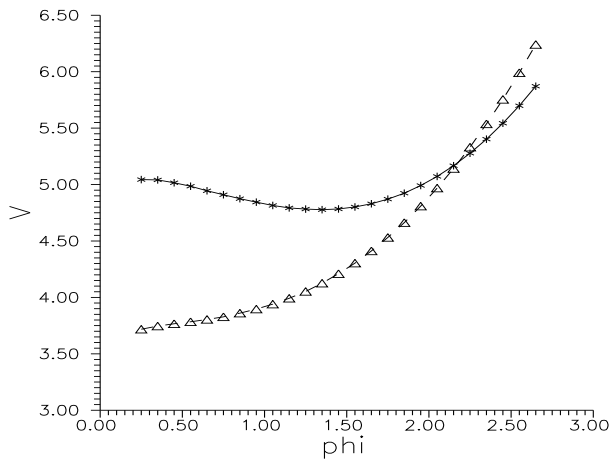,height=60mm,width=80mm,angle=0}
\caption{\label{fig.1_} The ultraviolet effective constraint potential at
$\gamma = 0.26$ (triangles, dashed line) and $\gamma = 0.262$ (crosses, solid
line); $\lambda =0.0025$ , $\beta = 12$. Error bars are about of the same size
as the symbols used. }
\end{center}
\end{figure}

In view of (\ref{UVPOT}) we use instead of the ultraviolet potential (\ref{uv})
the expression $U^{u-v}(\phi)=V^{u-v}(\phi)+ 3\, {\rm log}\, \phi$. (The term $
3\, {\rm log}\, \phi$ comes from the measure over the scalar doublet that has
$4$ real components.)

According to the results obtained in \cite{Z2010} at $\gamma \le \gamma_c$
there is the only minimum of the ultraviolet effective constraint potential
$U^{u-v}$. This is illustrated by Fig. \ref{fig.1_}. At the same time for
$\gamma > \gamma_{c}$ the ultraviolet potential has the only minimum at nonzero
value of $\phi$ (see also Fig. \ref{fig.1_}).

Our present numerical results for the ultraviolet potential show that at
$\gamma = 0.268$ the best fit to $U^{u-v}$ is
\begin{equation}
U^{u-v}(\phi) = const + 0.83(\phi - 2.75)^2
\end{equation}
At the same time the estimate given in Section 3 is (for $m_H =
4\sqrt{\frac{\gamma - \gamma_c^{(0)}}{\gamma}} \sim 1.1$):
\begin{equation}
U^{u-v}(\phi) = const + \frac{\gamma}{2G_{m_H}(0)}(\phi - v)^2 \sim const +
1.06 (\phi - 3.8)^2
\end{equation}

We observe the $20$ percent discrepance between the measured dispersion and its
tree level estimate and even larger discrepance between the value of $v$
calculated using the ultraviolet potential and its tree - level estimate. As
for the critical value $\gamma_c$, its tree level estimate is $\gamma_c^{(0)} =
0.24875$ while the fluctuational region is localized between $0.252$ and
$0.262$

\subsection{Infrared Effective Potential UDZ at  $\gamma_c$}

\begin{figure}
\begin{center}
 \epsfig{figure=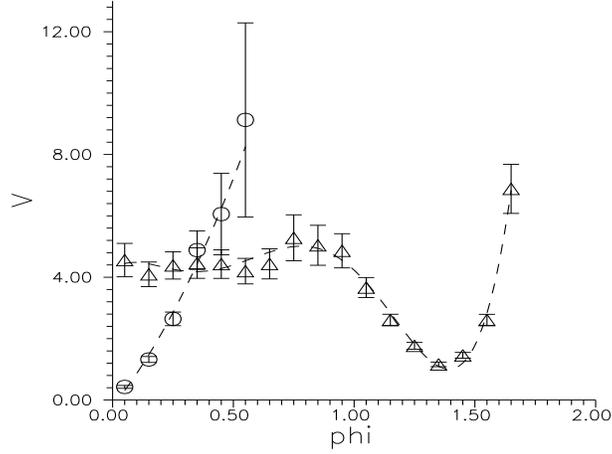,height=60mm,width=80mm,angle=0}
\caption{\label{fig.4} The infrared potential Eq.(\ref{ir}) UDZ  at $\gamma =
0.262$ (triangles), and  at $\gamma = 0.26$ (circles), $\lambda =0.0025$ ,
$\beta = 12$. }
\end{center}
\end{figure}

 This
potential has the only minimum at $\phi = 0$ for $\gamma \le \gamma_c=0.26$.
At $\gamma
> \gamma_{c}$  the UDZ potential has the nontrivial minima at nonzero
values of $\phi$ (see Fig. \ref{fig.4}). Already at $\gamma = 0.262 \sim
\gamma_{c2} $ there is the only nontrivial minimum of the potential that is
deeper than the other local minima observed.

\subsection{Infrared Effective potential UZ and the transition at $\gamma_c^{\prime}$}

\begin{figure}
\begin{center}
 \epsfig{figure=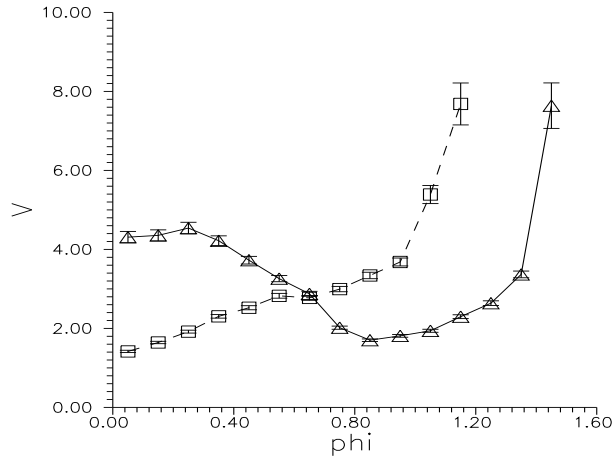,height=60mm,width=80mm,angle=0}
\caption{\label{uz} The infrared potential Eq.(\ref{ir}) UZ at $\gamma =
0.2575$ (squares), and $\gamma = 0.258$ (triangles);  $\lambda =0.0025$ ,
$\beta = 12$. }
\end{center}
\end{figure}

The infrared potential UZ  has the minimum at $\phi = 0$ for $\gamma \le
0.2575$ (see Fig.\ref{uz}).
 At $\gamma \ge 0.258$ the given
potential has nontrivial minimum at nonzero value of $\phi$ (Fig.\ref{uz}).

The given results point to $\gamma_{c}^{\prime}=0.25775\pm0.00025$ as to the
point, where the UZ potential changes its form.

\subsection{Z - boson mass from the infrared potential}

The present numerical results on the $Z$ - boson mass in the $Z$ - version of
Unitary gauge  confirm the results of \cite{Z2010}. Nonzero values of $Z$ -
boson mass are obtained at $\gamma > \gamma_c^{\prime}$. At the same time for
$\gamma < \gamma_c^{\prime}$ we observe large statistical errors for the $ZZ$
correlator. Therefore, in this region of the phase diagram the $Z$ - boson mass
cannot be calculated and we suppose it vanishes somewhere between $\gamma =
0.25$ and $\gamma = \gamma_c^{\prime}$.

At $\gamma = 0.268$ the infrared potentials give the value of $v = 2.95\pm
0.05$ that is to be compared with the mentioned above tree level estimate and
the value given by the ultraviolet potential. It is instructive to calculate
lattice $Z$ - boson mass using the given value of $v$ and the expression $m_Z =
v\sqrt{\frac{\gamma}{\beta \, {\rm cos}^2 \theta_W}}$. Using the value $v =
2.95\pm 0.05$ we obtain in this way $m^{\prime}_Z = 0.51\pm 0.01$. This is to
be compared with the value of lattice $Z$ - boson mass reported in
\cite{Z2010}: $m_Z = 0.49 \pm 0.01$. That's why at this value of $\gamma$ the
infrared potential gives reasonable estimate for the scalar field condensate.
At $\gamma = 0.262$ the infrared potential UDZ gives the value  $v_{UDZ} =
1.45\pm 0.05$ while the UZ potential gives $v_{UZ} = 2.35\pm 0.05$. These
values give $m^{\prime}_{UDZ} = 0.25\pm 0.01$ and $m^{\prime}_{UZ} = 0.40\pm
0.01$ while from \cite{Z2010} we get $m_Z = 0.29 \pm 0.01$. At the same time
expression $m_Z = v\sqrt{\frac{\gamma}{\beta \, {\rm cos}^2 \theta_W}}$ gives
zero lattice $Z$ - boson mass at $\gamma = \gamma_c$ (when $v$ is extracted
from the UDZ potential) or at $\gamma = \gamma^{\prime}_c$ (when $v$ is
extracted from the UZ potential). The value calculated in \cite{Z2010} differs
from zero at $\gamma_c$. At $\gamma_c^{\prime}$ we cannot calculate $m_Z$ due
to large statistical errors. Now we do not insist on the validity of one of the
mentioned estimates. Instead we suppose that within the Fluctuational region
the definition of the gauge boson masses becomes ambiguous because the usual
perturbation theory, most likely, does not work there.

\subsection{Renormalized fine structure constant}

In the present paper in order to calculate the renormalized fine structure
constant $\alpha_R = e^2/4\pi$ (where $e$ is the electric charge) we use the
correlator of Polyakov lines for the right-handed external leptons. These lines
are placed along the selected direction (called below imaginary "time"
direction). The space - like distance between the lines is denoted by $R$.
\begin{equation}
 {\cal C}(|\bar{x}-\bar{y}|)  =
 \langle {\rm Re} \,\Pi_{t} e^{2i\theta_{(\bar{x},t)(\bar{x},t+1)}}\,\Pi_{t} e^{-2i\theta_{(\bar{y},t)(\bar{y},t+1)}}\rangle.
\end{equation}
The potential is extracted from this correlator as follows
\begin{equation}
 {\cal V}(R) = -\frac{1}{L} { \rm log}\,  {\cal C}(R)
\end{equation}
Here $L$ is the size of the lattice in the imaginary "time" direction.

Due to exchange by virtual photons at large enough distances one would expect
the appearance of the Coulomb interaction
\begin{eqnarray}
 {\cal V}(r) & = & -\alpha_R \, {\cal U}_0(r)+ const,\,
\nonumber\\
{\cal U}_0(r) & = & -\frac{ \pi}{N^3}\sum_{\bar{p}\ne 0} \frac{e^{i p_3
r}}{{\rm sin}^2 p_1/2 + {\rm sin}^2 p_2/2 + {\rm sin}^2
 p_3/2}
 \label{V1}
\end{eqnarray}
Here $N$ is the lattice size, $p_i = \frac{2\pi}{L} k_i, k_i = 0, ..., L-1$.

However, at smaller distance the better fit to the potential is given by
\begin{eqnarray}
 {\cal V}(r) & = & -\alpha_R \, [{\cal U}_0(r)+\frac{1}{3}{\cal U}_{m_Z}(r)] +   const,\,
\nonumber\\
{\cal U}_m(r) & = & -\frac{ \pi}{N^3}\sum_{\bar{p}} \frac{e^{i p_3 r}}{{\rm
sin}^2 p_1/2 + {\rm sin}^2 p_2/2 + {\rm sin}^2
 p_3/2 +  {\rm sh}^2 m/2}
 \label{V2}
\end{eqnarray}
Here exchange by virtual massive $Z$ - bosons is taken into account.

We  substitute to (\ref{V2}) the  linear fit to the $Z$ - boson mass calculated
in \cite{Z2010}.

\begin{figure}
\begin{center}
 \epsfig{figure=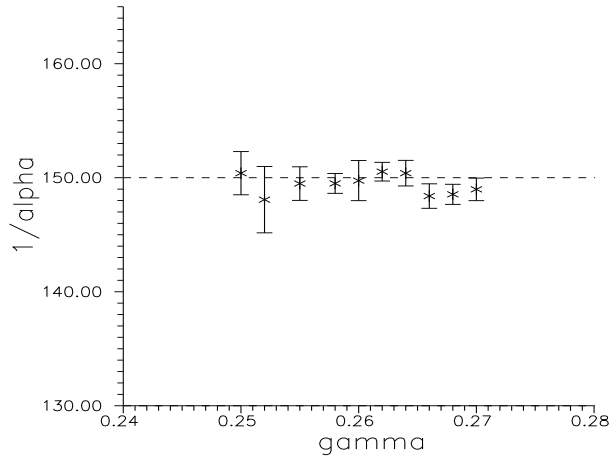,height=60mm,width=80mm,angle=0}
\caption{\label{fig.6} The renormalized fine structure constant as a function
of $\gamma$ at $\lambda =0.0025$ , $\beta = 12$.   }
\end{center}
\end{figure}

The results are presented in Fig. \ref{fig.6} and are to be compared with the
tree level estimate for the fine structure constant $\alpha^{(0)} \sim
\frac{1}{151}$ and the $1$ - loop approximation (when we assume bare value of
$\alpha$ to live at the scale $\sim 1$ TeV while the renormalized value lives
at the Electroweak scale $M_Z$):  $\alpha^{(1)}(M_Z/{1 \, {\rm TeV}}) \sim
\frac{1}{149.7}$.

 Fig. \ref{fig.6} demonstrates that the renormalized fine
structure constant calculated in the mentioned above way  is close to the one -
loop estimate (when the cutoff\footnote{For $\Lambda = 1000$ TeV, for example,
we would get the one - loop result $\alpha^{(1)}(M_Z/\Lambda) \sim 1/147$ that
seems to deviate already from our numerical results presented in
Fig.\ref{fig.6}.} $\Lambda$ in $\alpha^{(1)}(M_Z/\Lambda)$ is around $1$ TeV).
This confirms indirectly that the values of the $Z$ - boson mass calculated in
\cite{Z2010} are correct.

\section{Discussion}

In the present paper we have reported the results of our numerical
investigation of lattice Weinberg - Salam model at $\beta = 12$, $\lambda =
0.0025$, $\theta_W = 30^o$. For these values of couplings the bare Higgs boson
mass is close to $150$ GeV near to the transition between the Higgs phase and
the symmetric phase. All numerical simulations were performed on  rather small
lattices (of the size $8^3\times 16$). However, according to \cite{Z2010} the
most important results do not depend on the lattice size for the lattices of
the linear size up to $20$. Therefore, we feel this appropriate to publish our
results considered at the present moment as preliminary.

The effective potential in the model has been calculated in two different ways,
after two different gauge fixing procedures are applied. In both cases the
scalar doublet has the form $H = \left(\begin{array}{c}h\\0\end{array}\right)$,
where $h$ is real. However, the signs of $h$ are fixed differently. One of the
given effective potentials changes its form at $\gamma_c=0.26\pm 0.001$ (at
this point the ultraviolet potential calculated in \cite{Z2010} changes its
form as well), the other potential changes its form at
$\gamma_c^{\prime}=0.25775\pm 0.00025$. Such a pattern is typical for the
crossovers: different quantities change their behavior at different points on
the phase diagram.

At the present moment we imagine the pattern of the transition as follows. When
$\gamma$ is decreased $Z$ vortices become more and more dense. The first step
is the transition to the fluctuational region, where $Z$ - vortices and the
Nambu monopoles dominate. This occurs around $\gamma_{c2}\sim 0.262$ (see
\cite{Z2010}). This region may be to some extent similar to the mixed phase of
the second order superconductors. This phase of the superconductor appears when
the external magnetic field is present and the lattice of Abrikosov vortices
appears. These vortices are, in turn, the embryos of the normal phase within
the superconducting one. When the magnetic field achieves the second critical
value, these vortices overcome the repulsive forces and are transformed into
the homogenious symmetric phase. In our case the analogue of this phenomenon
may occur at the final step of the transition, where the lattice Z - boson mass
vanishes.

According to our results no signs of the two - state signal are found.
 It is worth mentioning that in classical second order
superconductors the transition to the normal phase is usually thought of as the
second order phase transition \cite{Landau}. Lattice simulations \cite{AHM3D},
in turn, indicate that the transition is a
 crossover in lattice Ginzburg - Landau Model (when it describes the second order superconductors)
 \footnote{In the 4D AHM, however, the situation is not so clear (see, for example,
\cite{AHM4D}).}. Our data, certainly, show that on the lattice $8^3\times 16$
the transition in the lattice Weinberg - Salam model is a crossover. This
follows from the fact that different observables change their behavior at
different points on the phase diagram. In particular, the infrared effective
potentials UDZ and DZ change their form at different points. At the same time
far from the transition both potentials give the same value of the scalar field
condensate. In \cite{Z2010} lattice masses were calculated for
$\gamma>\gamma_{c}^{\prime}$. At $\gamma<\gamma_{c}^{\prime}$ statistical
errors do not allow to estimate $Z$ - boson mass. Therefore, we do not exclude
at the present moment, that the second order phase transition may appear on the
lattices of larger sizes somewhere close to $\gamma_{c}^{\prime}$. At the same
time, it is very unlikely, that the first order phase transition may appear on
the larger lattices. To our opinion, the step - like change of the entropy
would manifest itself already on the lattice $8^3\times 16$ if it occurs on an
infinite lattice. In addition, we know that there is no first order phase
transition in Abelian Higgs Model at $M_H>M_Z$. In lattice Weinberg- Salam
model the crossover seems to us the preferred possibility.

The detailed analysis of the considered phase transition is, therefore, still
to be performed in order to understand its physics. In particular, it would be
very important to repeat our calculations on the larger lattices. Our present
results on the infrared effective potentials obtained on the lattice $8^3\times
16$ can be considered as a starting point of such an investigation.

It is also important to investigate more carefully the possible relationship
between the fluctuational region in the Electroweak theory and the mixed phase
of the superconductors. The possibility to approach continuum physics within
this region of the phase diagram is an important question as well. At a first
look in the fluctuational region the $Z$ - vortices and the Nambu monopoles
must contribute to the physical observables. This makes it impossible to use
only the conventional perturbation expansion around the trivial vacuum $h =
const$. The reason is that this expansion ignores the given topological objects
(at least, on the level of the the first terms of the loop expansion).
Therefore we suppose that
 this region cannot serve as a source
of the conventional continuum Electroweak physics due to the $Z$ - vortices and
Nambu monopoles that dominate there. It is worth mentioning, however, that our
calculation of the renormalized fine structure constant within this region
shows that the resulting values of $\alpha_R$ are surprisingly close to the one
- loop perturbative result $\alpha^{(1)}(M_Z/\Lambda)$ when the cutoff
$\Lambda$ is of the order of $1$ TeV.

Finally, we would like to notice that the  methods used (including the
calculation of the infrared effective potential) can be applied to the
investigation of the finite temperature Electroweak phase transition.

The authors kindly acknowledge discussions with V.I.Zakharov, V.A.Rubakov,
V.A.Novikov, and M.I.Vysotsky. This work was partly supported by RFBR grant
09-02-00338, 11-02-01227- , by Grant for leading scientific schools 679.2008.2.
This work was also supported by the Federal Special-Purpose Programme 'Cadres'
of the Russian Ministry of Science and Education. The numerical simulations
have been performed using the facilities of Moscow Joint Supercomputer Center,
and the supercomputer center of Moscow University.

\end{document}